\documentstyle[prb,preprint,aps]{revtex}

\begin{document}
\draft
\title{Eight-potential-well order-disorder ferroelectric model and 
effects of random fields}
\author{Zhi-Rong Liu and Bing-Lin Gu}
\address{Department of Physics, Tsinghua University,
Beijing 100084, People's Republic of China}
\author{Xiao-Wen Zhang}
\address{State Key Laboratory of New Ceramics and Fine Processing,
Department of Materials Science and Engineering, \\
Tsinghua University, Beijing 100084, People's Republic of China}
\maketitle

\begin{abstract}
An eight-potential-well order-disorder ferroelectric
model was presented and the phase transition was studied under the mean-field 
approximation. It was shown that the two-body interactions are able 
to account for the first-order and the second order phase 
transitions. With increasing the random fields in the system, a first-order 
phase transition is transformed into a second-order 
phase transition, and furthermore, a second-order phase transition is inhibited. 
However, proper random fields can promote the spontaneous appearance of a 
first-order phase transition by increasing the overcooled 
temperature. The connections of the model with relaxors were discussed.
\end{abstract}

%

\vspace{2mm}



Relaxor ferroelectrics (relaxors) experience no macroscopic 
phase transition at zero electric field.\cite{1} Nevertheless, a ferroelectric 
phase can be induced by an external electric field at low temperatures. For 
example, in Pb(Mg$_{1/3}$Nb$_{2/3})$O$_3$ (PMN), a first-order induced phase 
transition, from the pseudo cubic phase to a macroscopically polar phase, 
has been disclosed by Ye {\it et al.}.\cite{2} A field-temperature 
phase diagram was proposed for PMN and a minimum threshold field 
strength $E_{th}$ to induce the ferroelectric phase 
was determined.\cite{2} The random interactions between polar microregions 
and the random electric fields existing in the system 
were usually adopted to interpret the 
special characteristics of relaxors.\cite{3,4,5} Glinchuk {\it et al.} showed 
that strong random fields can make the long-range order parameter 
disappear, and thus qualitatively explained the absence of the zero-field 
phase transition in relaxors.\cite{6} However, they only considered 
a second-order phase transition and no external electric field 
was involved. It appears, therefore, necessary to investigate the 
effects of random fields and external fields on the first-order 
phase transition.

A pseudospin model\cite{7} is usually used to interpret the phase-transition 
phenomena of order-disorder ferroelectrics in which the active ions have two 
equilibrium sites. In order to explain the first-order phase transition, 
four-spin coupling terms should be introduced and the 
four-body coupling strength is required to be larger than the 
two-body one.\cite{8} On the other hand, PMN presents 
a disordered pseudo cubic structure with important atomic shifts around 
the special positions of the ideal perovskite structure,\cite{9} i.e., 
the polar ions in PMN have more than two equilibrium positions. 
Therefore, a multi-direction ``spin'' system should be studied in order 
to understand the special properties of PMN.

For the reasons mentioned above, we present an eight-potential-well ferroelectric 
model and investigate the the phase transition under various condictions 
in this paper.


For PMN, the electric field-induced macropolar phase 
has trigonal symmetry with point group $3m$,\cite{2} which suggests 
that for each polar ion there are eight potential wells along 
\{111\}-equivalent directions. When an external field 
is applied along one of eight \{111\}-equivalent directions, the Hamiltonian 
matrix of the order-disorder system with eight potential wells is assumed to be:
\begin{equation}
H=\left[
\begin{array}{cccccccc}
-Ep_0 & \frac{\Omega}{2} & 0 & 0 & 0  & 0 & 0 & 0\\
\frac{\Omega}{2} & Ep_0 & 0 & 0 & 0  & 0 & 0 & 0\\
0 & 0 & -\frac{1}{3}Ep_0 & \frac{\Omega}{2} & 0 & 0 & 0  & 0 \\
0 & 0 & \frac{\Omega}{2} & \frac{1}{3}Ep_0 & 0 & 0 & 0  & 0 \\
0 & 0 & 0 & 0 & -\frac{1}{3}Ep_0 & \frac{\Omega}{2} & 0 & 0 \\
0 & 0 & 0 & 0 & \frac{\Omega}{2} & \frac{1}{3}Ep_0 & 0 & 0 \\
0 & 0 & 0 & 0 & 0 & 0 & -\frac{1}{3}Ep_0 & \frac{\Omega}{2} \\
0 & 0 & 0 & 0 & 0 & 0 & \frac{\Omega}{2} & \frac{1}{3}Ep_0  
\end{array} \right],
\end{equation}
where $E$ is the electric field strength, and $p_0$ is 
the magnitude of the dipole when an ion locates in 
a certain well. A factor 1/3 is introduced in terms of 
Eq. (1) when the directions of the well are not parallel to that of 
the external field. $\Omega$ is the tunneling frequency. 
Only the tunneling between wells with opposite directions are considered 
here. In ferroelectric systems, 
polar ions interact with each other and with different kinds of defects in the system. 
Under a mean field 
approximation, the interactions upon a certain ion may be represented 
by an equivalent field, i.e., 
\begin{equation}
Ep_0=J\frac{\langle p \rangle}{p_0}+E_{ext}p_0+E_{rand}p_0,
\end{equation}
where $\langle p \rangle$ is the thermal average value of dipole moments, and 
$J$ is the coupling energy. Only two-body interactions 
are involved here because a linear relation is assumed between $\langle p \rangle$ 
and $E$. $E_{ext}$ is the applied external electric 
field. $E_{rand}$ is the internal random fields in the system 
which come from the direct interactions of dipoles, fields 
created by point charge defects and atomic ordering, etc.. 
$E_{rand}$ is assumed to 
have a Gaussian distribution:
\begin{equation}
\rho(E_{rand})=\frac{1}{\sqrt{2\pi}\sigma_e}
\exp\left[\frac{-E_{rand}^2}{2\sigma_e^2}\right],
\end{equation}
where $\sigma_e$ is the distribution width. By using Eqns. (1)-(3), the 
order parameter $\langle p \rangle$ can be determined for any temperature.


At first, the phase transition temperatures are calculated when 
there is no external field and random fields, which are shown 
in Fig. 1. It can be seen that the long-range order can appear 
only at $\Omega\leq 2J$. If the tunneling movement does not exist ($\Omega=0$), 
the critical temperature is $T_c=\frac{J}{3k_B}$ where $k_B$ is the 
Boltzmann constant. An interesting characteristic of Fig. 1 is that 
both the second-order phase transition and the first-order phase transition 
exist for the present model. It is quite different from the 
case of two-direction pseudospin model\cite{7} where no 
first-order phase transition exists if the number of coupled spins 
is two. In Fig.1, there is a tricritical point between the 
second- and the first-order phase transitions, which is determined 
as $\Omega_{tri}=0.56J$ and $T_{tri}=0.23J/k_B$. For a first-order 
phase transition, the overcooled temperature $T_-$ decreases rapidly 
with increasing $\Omega$ and vanishes at $\Omega=\frac{2}{3}J$. When 
$\frac{2}{3}J<\Omega<2J$, the long-range order cannot spontaneously 
appear in a cooling process, while a polar state can be induced 
at low temperatures. This is qualitatively consistent with the 
behavior of PMN.\cite{2}

Solid lines in Fig. 2 depict the phase transition temperatures 
versus external field when $\Omega=J$ and $\sigma_e=0$. 
At small external fields, the phase transition 
is the first-order type. And at large external fields, a second-order 
phase transition occurs. 
A tricritical point also appears here. For a first-order 
phase transition, the induced ferroelectric phase 
disappeares at the overheated temperature in a heating process. 
It is seen from the figure that the overheated temperature increases with 
increasing the field strength, whereas the zero-field depoling of 
an induced state occurs always at the temperature $T_{do}$ 
(see Fig. 2), independently of the initial poling field. The existence 
of $T_{do}$ and the tricritical point agrees with the experiments in 
PMN.\cite{2,10} The curve of the overcooled temperature can also 
be interpreted as the curve of the threshold field 
strength $E_{th}$ to induce a ferroelectric state at different 
temperatures. $E_{th}$ decreases with decreasing temperature in our 
calculation. In experiments,\cite{2} however, $E_{th}$ decreases firstly 
and then increases with decreasing temperature. The discrepancies 
come from the dynamical process, which is not considered in the 
calculation. At low temperatures, the 
system is frozen into a glass-like state.\cite{3,4} The 
dynamical process may be so slow that the induced polar phase cannot 
come into being even if the polar phase has smaller free energy. 
As a result, $E_{th}$ increases at low temperatures ($T<T_{do}$). 

In Fig. 2, when the external 
field is larger than the tricritical point, there 
is no ``real'' phase transition in the sense that the spontaneous 
breakdown of symmetry does not occur. Nevertheless, the curves of 
the average polarization $\langle p \rangle$ in cooling processes 
(see Fig. 3) indicate that $\langle p \rangle$ increases rapidly 
in a small temperature range but changes slowly at other temperatures 
when the external field is a little larger than the tricritical value 
(curve c in Fig. 3). Thus, a second-order ``phase transition'' 
may be approximately defined at the temperature where $\langle p \rangle$ changes 
most quickly. When the external field is much larger than the 
tricritical value (curve d in Fig. 3), $\langle p \rangle$ varies 
rather smoothly, and the phase transition is difficult 
to determine. The features of the polarization curves in 
Fig. 3 are consistent with experiments.\cite{11}

To explore the effects of random fields, the phase transition 
temperatures for the random fields $\sigma_e=0.2J/p_0$ are also depicted 
in Fig.2 (dashed lines). Compared with the curves with no random field 
(solid lines), the critical temperature and the overheated temperature 
decrease, and the range of the first-order phase transition gets smaller, 
i.e., the phase transition is compressed. It should be noted 
that the overcooled temperature increases when random fields exist. 
Thus the phase transition is promoted by the random fields in the sense that 
a spontaneous phase transition is easier to occur, which is in conflict
with the common sense.

The continuous effects of random fields on the phase transition 
temperatures are shown in Fig. 4 for $\Omega=J$. By enhancing 
random fields, the interval between the overheated 
and the overcooled temperatures shrinks 
gradually, and vanishes at a tricritical point. 
Beyond the tricritical point, the random fields make the 
critical temperature decrease, and finally inhibit the phase 
transition completely. The promotion effect 
of random fields on the spontaneous first-order phase transition 
is obvious in Fig. 4: the critical temperature keeps approximately 
constant while the overcooled temperature increases extraordinarily. 
This unusual effect seems to challenge the previous idea that 
the phase transition in PMN is inhibited by strong random 
fields. Perhaps the actual situation is that the random fields are 
not strong enough to promote a spontaneous phase transition.

At last, a phase diagram on the $\Omega-\sigma_e$ plane is given 
in Fig. 5. The $\Omega-\sigma_e$ plane is divided into three regions 
where a first-order phase transition, a second-order phase transition 
and no phase transition can occur, respectively. The boundary line 
between the first-order and the second-order phase transition regions 
is a tricritical line that is composed of tricritical points.


In summary, the phase transition of an eight-potential-well order-disorder 
ferroelectric system was investigated in this paper. The effects of the 
couplings, the external fields, and the internal random fields are discussed. 
The connections of the model with PMN were discussed.


This work was supported by the Chinese National Science Foundation 
and State Key Program of Basic Research 
Development, 
and the authors would like to thank Dr. Jian-She Liu for helpful 
discussions.

\begin{figure}[tbp]
\caption{Phase transition temperatures as functions of tunneling frequency with 
no random fields. The solid
line, the dashed line, and the dotted line represent the critical, the 
overheated, and the overcooled temperatures, respectively. The temperatures 
and the tunneling frequency are measured in units of $J/k_B$ and $J$ 
respectively. }
\end{figure}

\begin{figure}[tbp]
\caption{Phase transition temperatures as functions of external field when 
$\Omega=J$. The 
lines above tricritical points represent critical temperatures. Below 
tricritical points, lines on the left represent overcooled temperatures, and 
that on the right represent overheated temperatures. The solid and dashed lines 
correspond to the case of $\sigma_e=0$ and $0.2J/p_0$, respectively. External 
fields are measured in unit of $J/p_0$ and temperatures in unit of $J/k_B$. }
\end{figure}

\begin{figure}[tbp]
\caption{Average polarization $\langle p \rangle$ (in unit of $p_0$) as function of temperature 
(in unit of $J/k_B$) in cooling processes at $\Omega=J$ and $\sigma_e=0$. 
Curves a, b, c, and d correspond to 
the external field $E_{ext}=0.075$, 0.095, 0.115, and 0.15 $J/p_0$, respectively. 
 }
\end{figure}

\begin{figure}[tbp]
\caption{Phase transition temperatures (in unit of $J/k_B$) versus random 
fields width (in unit of $J/p_0$) when $\Omega=J$. The solid, dashed and 
dotted lines represent the critical, overheated, and overcooled 
temperatures, respectively.}
\end{figure}

\begin{figure}[tbp]
\caption{Phase diagram on the $\Omega-\sigma_e$ plane. $\alpha$, $\beta$, and 
$\gamma$ designate the regions where the first-order, the second-order, and 
no phase transition can occur. $\Omega$ and $\sigma_e$ are measured in units 
of $J$ and $J/p_0$, respectively. }
\end{figure}


\vspace{2mm}



\begin{references}
\bibitem{1} L. E. Cross, Ferroelectrics {\bf 76}, 241 (1987).

\bibitem{2}  Z. G. Ye and H. Schmid, Ferroelectrics {\bf 145}, 83 (1993).

\bibitem{3}  D. Viehland, S. J. Jang, and L. E. Cross, J. Appl. Phys. {\bf 68%
}, 2916(1990).

\bibitem{4}H. Gui, B. L. Gu, and X. W. Zhang, Phys. Rev. B {\bf 52}, 
    3135 (1995).

\bibitem{5}  V. Westphal, W. Kleemann, and M. D. Glinchuk, Phys. Rev. 
Lett. {\bf 68}, 847 (1992).

\bibitem{6}  M. D. Glinchuk and V. A. Stephanovich, J. Phys.: Condens. Matter 
{\bf 6}, 6317 (1994).

\bibitem{7}  R. Blinc and B.Zeks, {\it Soft Modes in Ferroelectrics and 
Antiferroelectrics} (North-Hollad, Amsterdam, 1974).

\bibitem{8}  C. L. Wang, Z. K. Qin, and D. L. Lin, Phys. Rev. B {\bf 40}, 
680 (1989).

\bibitem{9} N. de Mathan, E. Husson, G. Calvarin, J. R. Gavarri, 
A. W. Hewat, and A. Morell, J. Phys.: Condens. Matter {\bf 3}, 8159 (1991).

\bibitem{10} G. Calvarin E. Husson, and Z. G. Ye, Ferroelectrics {\bf 165}, 
349 (1995).

\bibitem{11} Z. G. Ye, Key Engineering Materials {\bf 155-156}, 81 (1998).

\end{references}
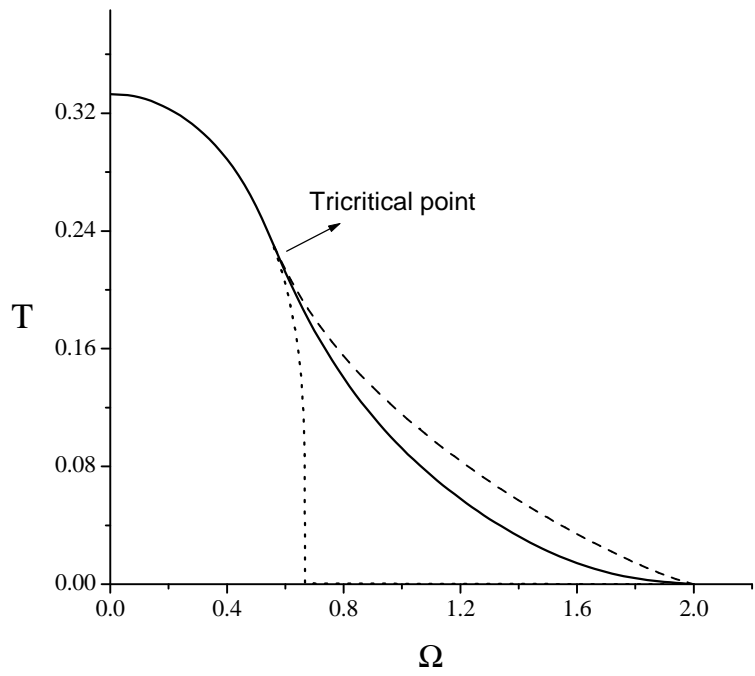
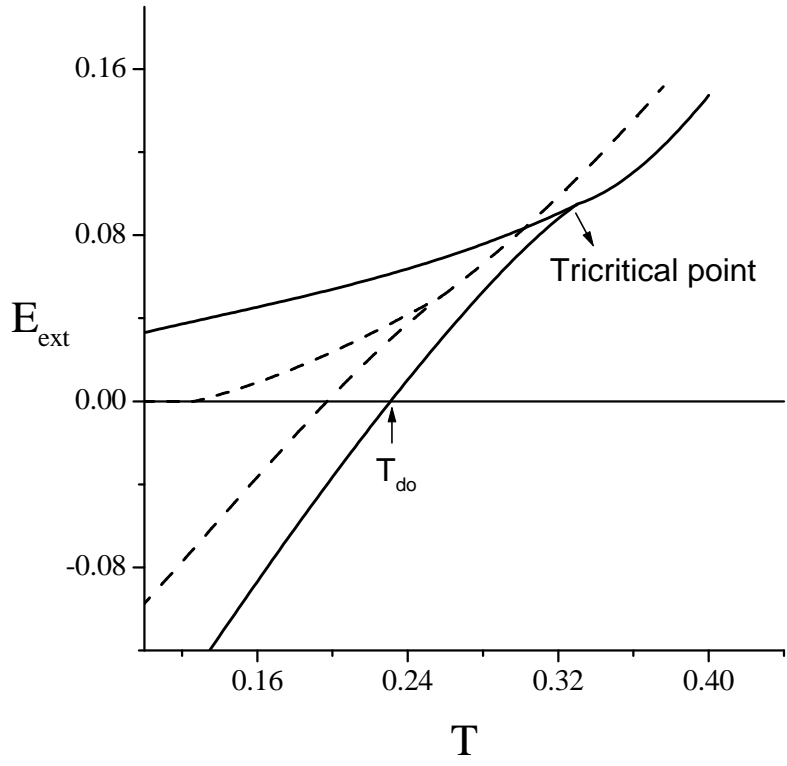
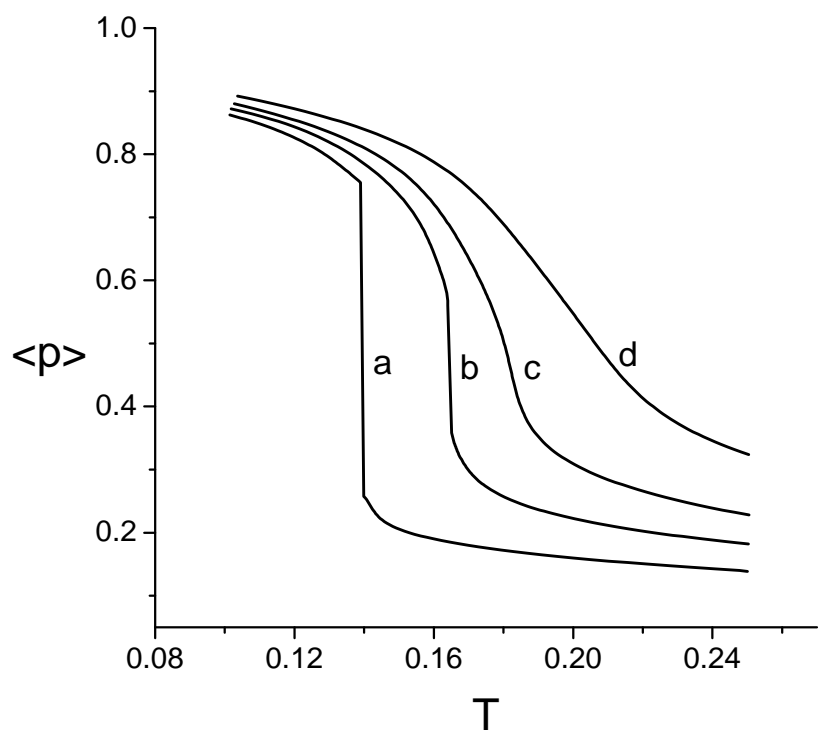
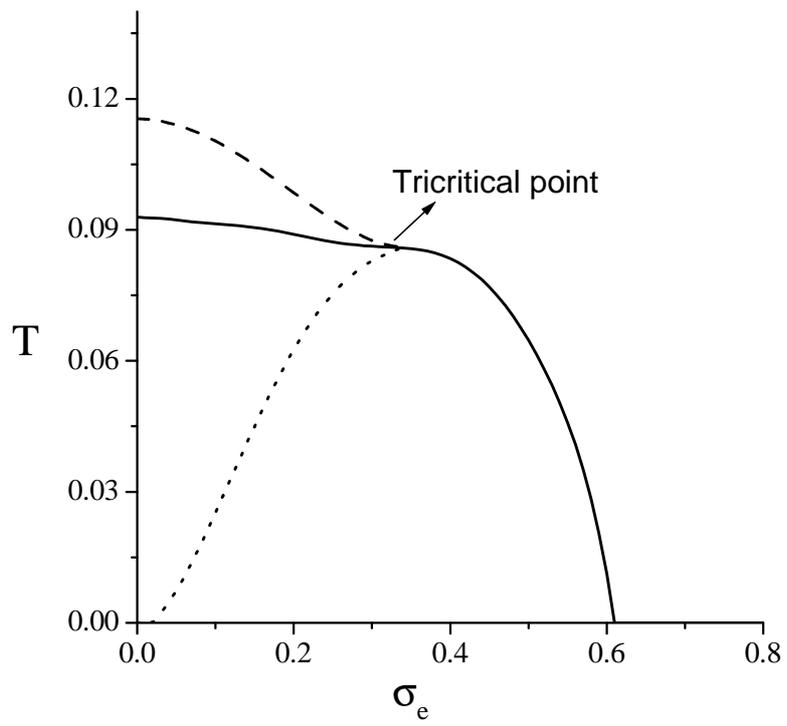
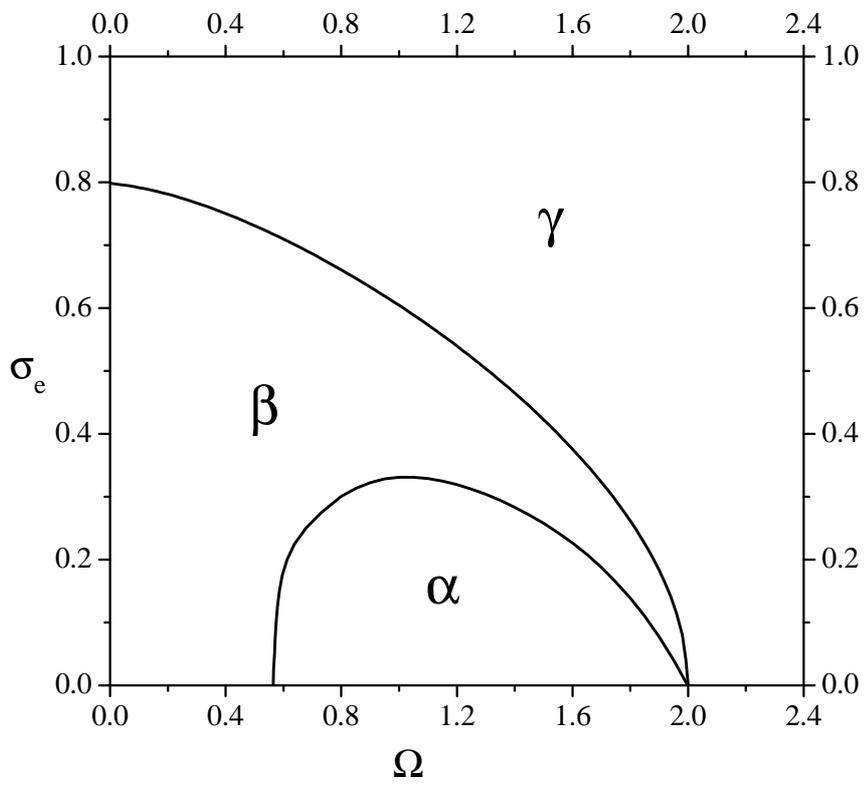
\end{document}